\documentclass[prb,twocolumn,showpacs,superscriptaddress]{revtex4-1}                    

\usepackage{graphicx,amssymb,amsmath,color,bbm}

\begin{document}


\newcommand{\dirop}{\mathcal{D}}



\title{Lattice generalization of the  Dirac equation to general spin and the
  role of the flat band} 

\author{Bal\'azs D\'ora}

\email{dora@pks.mpg.de}

\affiliation{Department of Physics, Budapest University of Technology and
  Economics, Budafoki \'ut 8, 1111 Budapest, Hungary} 

\author{Janik Kailasvuori}

\affiliation{Max-Planck-Institut f\"ur Physik komplexer Systeme, N\"othnitzer
  Str. 38, 01187 Dresden, Germany} 
\affiliation{International Institute of Physics, Universidade Federal do Rio Grande do Norte, 59078-400 Natal-RN, Brazil}

\author{R. Moessner}

\affiliation{Max-Planck-Institut f\"ur Physik komplexer Systeme, N\"othnitzer
  Str. 38, 01187 Dresden, Germany} 

\date{\today}

\begin{abstract}

We provide a novel setup for generalizing the two-dimensional pseudospin $S=1/2$ Dirac equation, arising in graphene's honeycomb lattice, to general pseudospin-$S$. We engineer these band
structures as a nearest-neighbor hopping Hamiltonian involving stacked
triangular lattices. We obtain multi-layered low energy excitations around
half-filling described by a two-dimensional Dirac equation of the form 
$H=v_F\bf S\cdot p$, where $\bf S$ represents an arbitrary spin-$S$ (integer or
half-integer). For integer-$S$, a flat band appears, whose presence modifies
qualitatively the response of the system.  Among physical observables, the density 
of states, the optical conductivity and the peculiarities of Klein tunneling are investigated. 
We also study Chern numbers as well as the zero-energy Landau level degeneracy. 
By changing the stacking pattern, the topological properties are altered significantly, with 
no obvious analogue in multilayer graphene stacks.

\end{abstract}


\maketitle

\section{Introduction}

Since the seminal work on monolayer graphene, a single sheet of carbon atoms
forming a honeycomb lattice, in 2004\cite{novoselov1}, a lot of attention has
been focused on this material. 
Its low energy properties close to half filling (i.e. pristine graphene) are
well described by a two-dimensional massless Dirac equation, with the speed of
light replaced by the appropriate Fermi velocity $\sim 10^6$~m/s. Most of the
unusual electronic properties of this material can be traced back to the
massless Dirac nature of its quasiparticles and their unusual Berry phase.
These include its linearly vanishing density of states (DOS) around half filling
resembling a d-wave superconductor, unusual Landau quantization in a
perpendicular magnetic field and the anomalous half-integer quantum Hall effect.
Additionally, phenomena such as the universal optical conductivity and high optical transparency, Klein
tunneling through electric barriers are also distinguishing features.

The appearance of the massless Dirac equation has triggered further research to
find out whether other systems can possess similar behaviour or even
generalizations of the $S=1/2$ Dirac physics to e.g. higher dimensions,
including additional terms.  In the context of ultracold atom in optical lattices, several 
proposals have
been put forward to realize a generalization of graphene physics in terms of
the $S=1$ Dirac equation\cite{bercioux,shen,greens1,apaja}.
Generalizations to  higher $S$ with spin-dependent hoppings\cite{lan} as well as with artificial magnetic field\cite{kennett}
are also possible.

Here, we present a family of lattices  whose low energy excitations 
around given fillings are described by a generalized two-dimensional Dirac
equation, 
\begin{gather}
H=v_F{\bf S \cdot p},
\label{hamilton}
\end{gather}
where ${\bf p}=(p_x,p_y,0)$, and ${\bf S} = (S_x, S_y, S_z)$ is the matrix
representation of an arbitrary spin $S$ (integer or half integer), and $v_F$ is  
the Fermi velocity. These lattices consist of stackings of triangular layers,
and include slabs of face-centred cubic and hexagonal close-packed lattices as
special cases.
Technically, the notion Weyl Hamiltonian\cite{lan} is more appropriate for Eq. \eqref{hamilton} for $S>1/2$, though we 
 refer to it as generalized Dirac equation (sometimes omitting the "generalized") since our motivation comes primarily from graphene and its pseudospin-1/2
Dirac equation.

Our model is characterized by considerable simplicity and tunability. Furthermore, it contains a unique feature, absent from 
previous lattice realizations of higher spin $S$ Hamiltonians: the possibility 
of---through a simple lateral shift in the layer positions---changing the chiral properties of individual interlayer hoppings, without 
changing the spectrum. However, there are considerable
changes on other properties, for example topological properties such as the multiplicity of the zero-energy Landau level degeneracy
in magnetic field. 

The paper is organized as follows. In Sec. II we introduce the lattice and discuss some of its general features.  We then proceed to analyze the properties resulting from such a band structure:
density of states (Sec. III.), optical conductivity (Sec. IV.),  
Chern numbers and spin Chern numbers of the band structure (Sec. V.), 
zero mode degeneracy in both a uniform as well as a nonuniform magnetic field (Sec. VI.)  and Klein tunneling (Sec. VII.) for the spin-1 case, 
focusing on tunneling into the flat band. 
We also derive the general matching condition for the wavefunction for arbitrary pseudospin-$S$.
Finally, the relevant symmetry properties are highlighted in an appendix.

The $S=1/2$ version
is realized in graphene\cite{castro07} and on the surface of 3D topological
insulators\cite{hasankane}. Of the recent proposals for the $S=1$ case\cite{bercioux,shen,greens1,apaja}, our construction includes the dice lattice.
Our work in many respect is complementary to Ref. \onlinecite{lan}, where diverse properties (topology, transport) of the spin-$S$ Dirac equation were studied using a different
lattice realization with spin-dependent hoppings.

\section{Band-structure engineering}

To set the stage, let us first cast the analysis of graphene's honeycomb
lattice in a form that lends itself to generalization. The bipartite
honeycomb lattice has two atoms (A and B) per unit cell; each sublattice
forms a triangular lattice and the hopping Hamiltonian in
Fourier space takes the form
\begin{gather}
H = \left[
\begin{array}{cc}
0  & tf({\bf k})\\
tf^*({\bf k}) & 0
\end{array}
\right],
\end{gather}
where $t$ is the hopping amplitude and $a$ the intercarbon distance, while
$f({\bf k})=1+2\exp(i3k_ya/2)\cos(\sqrt{3}k_xa/2)$.   

At half-filling the Fermi surface consists of two inequivalent Dirac 
points  $K$ and $K^{\prime}$ at momenta $ \pm (2\pi/3\sqrt{3}a,2\pi/3a)={\bf k}_\pm$, respectively. Expanding around 
these points leads to two copies of the two-dimensional Dirac equation for $S=1/2$ with the 
sublattice providing  the (pseudo)spin degree of freedom:
\begin{gather}
S_x=\frac{1}{2}\left(\begin{array}{cc}
0 & 1\\
1 & 0
\end{array}
\right),
\hspace*{6mm}
S_y=\frac{1}{2}\left(
\begin{array}{cc}
0 & -i\\
i & 0 
\end{array}\right)
\end{gather}
$v_F=3ta$, the missing factor $1/2$ as opposed to graphene\cite{castro07}
arising since the eigenvalues of the spin are $\pm 1/2$.

\begin{figure}[h!]

 a.)
{\includegraphics[width=8cm]{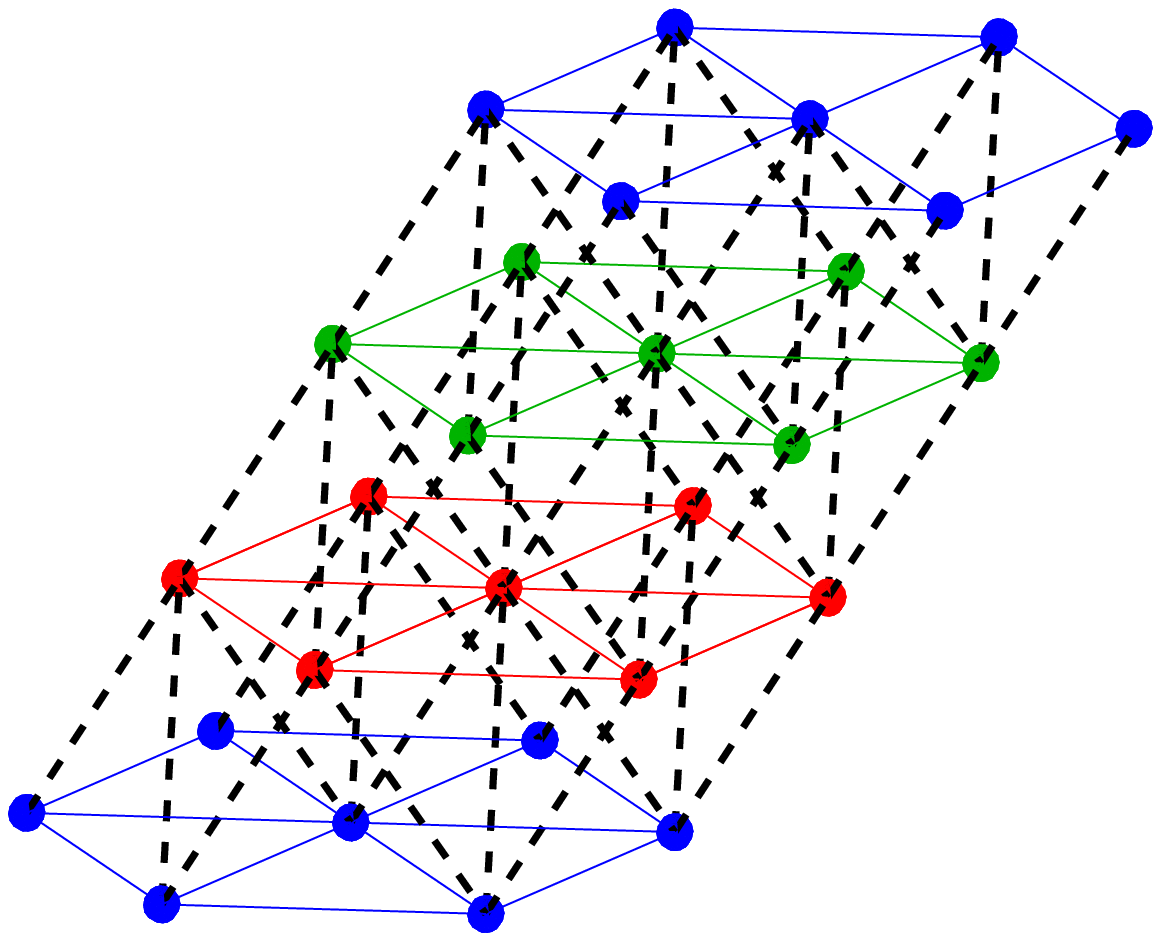}}
\bigskip

b.)
{\includegraphics[width=6cm]{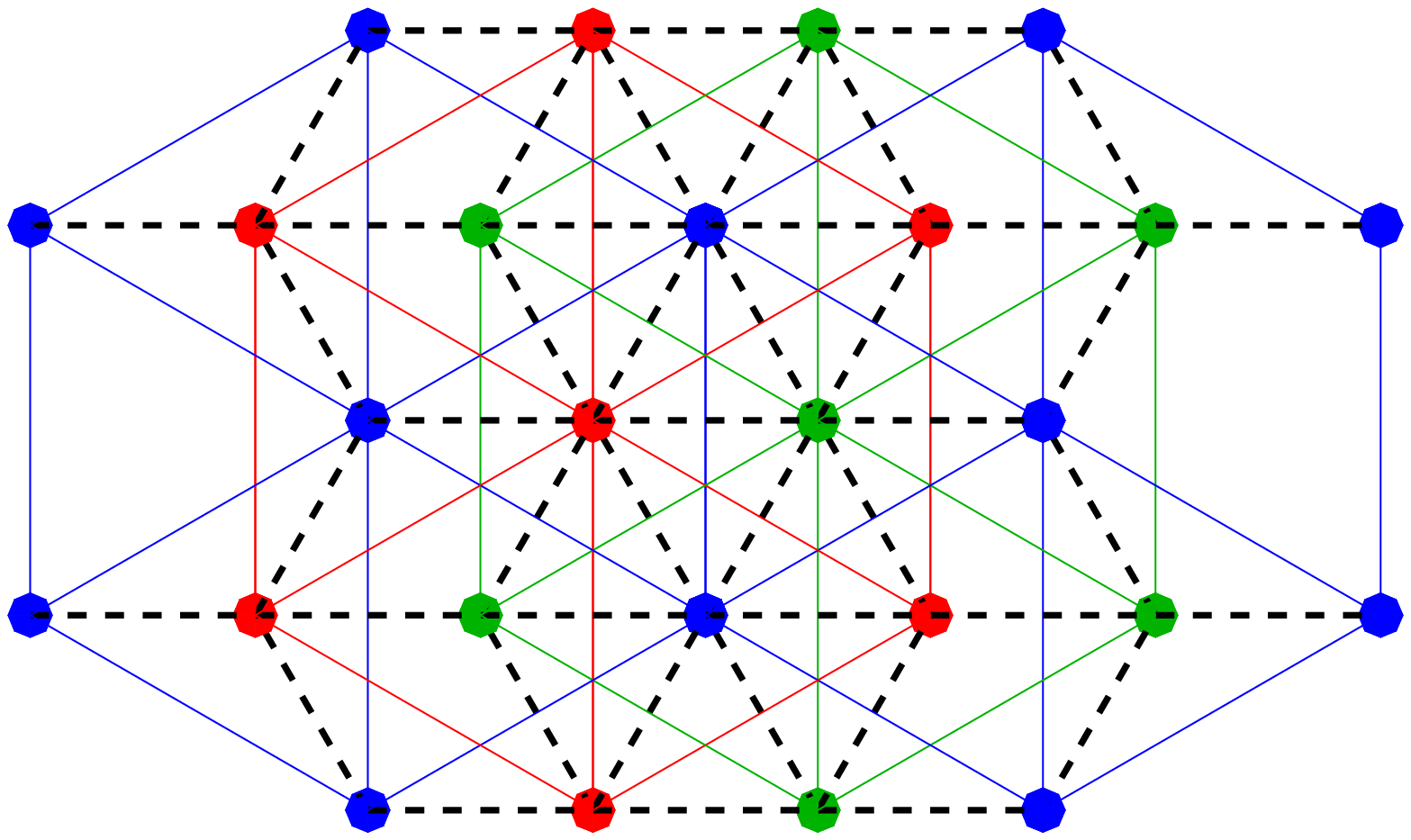}}

\caption{(Color online) The schematic representation of the  family of lattice
  models leading to the spin-$S$ Dirac equation is shown from the side (a) and from above (b). The dashed lines
  denote interlayer hopping processes, while the intralayer thin solid lines
  are guide to the eye, emphasizing the planar triangular structure, but do not
  represent any hoppings. The lowest and highest (blue A) planes are exactly on top
  of each other. For $S=1/2$, only two adjacent layers need to be considered
  (e.g. red A, and green B), for $S=1$, three neighbouring layers (e.g. lower blue A,
  red B and green C), for $S=3/2$, all four layers, while for higher $S$'s, one
  needs to continue up- or downwards with ABCABC... stackings. Note
  that the interlayer hoppings should be unequal to realise the perfect Dirac
  equation, Eq. \eqref{seq32}}.  
\label{sp12}
\end{figure}

If we now think of the honeycomb lattice as a layered structure, with the A
and B triangular sublattices  offset in height by an amount $h$, it is
natural to ask what happens if one adds a third, and then further, triangular
layers (see Fig. \ref{sp12}). For small $h$, the nearest neighbours of a given
site are in the layers directly above and below. When the layers are stacked in the
sequence of the face-centred cubic lattice in a [111] direction, one obtains
a band diagonal hopping Hamiltonian for a system of $2S + 1$ layers: 

\begin{widetext}
\begin{gather}
H_S = t \left[
\begin{array}{cccc}
0 & \alpha_{01} f ({\bf k}) &  0 & \\[1ex]
\alpha^*_{01} f^*({\bf k}) & 0 &  {\alpha_{12} f ({\bf k})}_{\ddots} &  0 \\[1ex]
0 & \alpha^*_{12} f^* ({\bf k})_{\ddots} & 0_{\ddots} & \alpha_{2S-1, 2S} f
({\bf k}) \\[1ex] 
& 0 & \alpha^*_{2S-1, 2S} f^* ({\bf k}) & 0 
\end{array}
\right],
\label{eq:04}
\end{gather}
\end{widetext} 
where we have  allowed for different interlayer hopping strengths by
introducing the $\alpha$'s. Indeed, regardless of the choice of $\alpha_{i,
  i+1}$, several properties of the spectrum of ${H_S}$ immediately follow
from the form of its characteristic polynomial $C_S (\lambda, {\bf k}) =$ det
$(H_S - \lambda \mathbbm{1})$ which reads: 

\begin{gather}
C_S  (\lambda, {\bf k}) = \left|t f ({\bf k}) \right|^{2S+1} {\rm det} \left[
  (\alpha^+ + \alpha^-) - \tilde{\lambda} \mathbbm{1} \right] 
\end{gather}
where $\tilde{\lambda} = \lambda/ |f({\bf k})|$ is independent of ${\bf k}$,
and so is $\alpha^+_{ij} = (\alpha^-_{ji})^* = \alpha_{i, i+1} \delta_{j, i+1}$.
Firstly, near the Dirac points  ($K$ and $K^\prime$), all bands are linearly dispersing, simply because
$|f({\bf k})| \propto |{\bf k}-{\bf k}_\pm|$. Secondly, for integer $S$  
the Hamiltonian must display a flat band.  The matrix $\alpha^++\alpha^-$  possesses a  symmetric spectrum (if $\tilde{\lambda}$ is an eigenvalue, then
 so is also $-\tilde{\lambda}$).  The Hamiltonian also possesses this symmetry, which can be phrased 
 as a chiral symmetry 
 $\Sigma H({\bf k})\Sigma^\dagger =-H({\bf k})$ with $\Sigma$ a unitary operator, as further discussed in the appendix~\ref{s:symmetry}. 
   For an odd number $2S+1$ of bands the chiral symmetry implies that 
(at least) one eigenvalue $\tilde\lambda$ must be zero, 
which translates into a flat band $\lambda = |f (k)| \tilde{\lambda} =0.$
~\footnote{This ties in with the observation of Ref. \onlinecite{greens1} that
 flat bands go along with integer spin Berry's phases.}

If now, in addition, we choose the interplane hopping amplitudes so that
$\alpha^+ = S^+$, where $S^+ = S_x + iS_y$ is the raising operator for spin
$S$, we obtain a spectrum $E_n ({\bf k}) = n t | f ({\bf k}) |$, where $n = -S,
-S+1, ..., S$. This requires placing the adjacent layers at certain distances from each other, 
so that the overlap of the wavefunctions would produce the appropriate hopping integrals between subsequent layers,
whose relative strength is further specified
in Eq. \eqref{spinmatrix}.

As a result, we obtain an effective Hamiltonian near the $K$ point,
\begin{gather}
H_S ({\bf p}) = v_F  {\bf S} \cdot {\bf p}.
\label{eq:spinham}
\end{gather} 
Here ${\bf p}={\bf k}-{\bf k}_+$ measures the (small) distance from the Dirac point at $K$, and similarly for the $K'$ point.
If we finally add a
potential of strength $\Delta S_z$, which can in principle be generated
straightforwardly via an electric field applied perpendicular to the layers,
representing distinct chemical potentials for each layer, we have
\begin{gather}
E_n ({\bf p}) = n \sqrt{v^2_F \left( p^2_x + p^2_y \right) + \Delta^2} ~~.
\label{eq:invariably}
\end{gather} 
For integer $S$, $n = 0$ invariably corresponds to a flat
band, however, no longer due to chiral symmetry but due to a less general symmetry that is specific to the low-energy Dirac-like 
Hamiltonians and that requires fine-tuning of the parameters $\alpha_{ij}$. Fortunately for experimental 
realizations, rather natural setups like equidistant layers will satisfy the conditions for a flat band, as discussed in the appendix~\ref{s:symmetry} in more detail.

It can now be verified straightforwardly that the cases $S = 1/2$ and $S = 1$
correspond to the known instances of the honeycomb and dice (or $T_3$)
lattices{\cite{castro07,bercioux}, respectively. The $S = 3/2$ case for four
  layers reads 
 \begin{gather}
H=\left[
\begin{array}{cccc}
0  & \sqrt{3} tf({\bf k}) & 0 & 0 \\
\sqrt{3} tf^*({\bf k}) & 0 &  2tf({\bf k}) & 0 \\
0 & 2tf^*({\bf k}) & 0 & \sqrt{3}tf({\bf k})\\
0& 0& \sqrt{3}tf^*({\bf k}) &0 \end{array}
\right],
\label{seq32}
\end{gather}
which can be supplemented with an additional gap, coming from $\Delta S_z$ with
$S_z=\textmd{diag}(3/2,1/2,-1/2,-3/2)$. A very similar lattice structure has been proposed in Ref. \onlinecite{watanabe}.

Note that the simple form of the Hamiltonian also gives immediate access to the
wavefunctions in layer (pseudospin) space, as its eigenfunctions are obtained from a simple 
rotation in spin space: the quantization axis of ${\bf S}$ is given by an effective field
direction, ${\bf h}$, whose components are given by  Re$f ({\bf k})$, Im$f ({\bf
  k})$ and $\Delta$, respectively. The various $\alpha$-prefactors above and below the
diagonal are chosen according to the conventional matrix representation of the
spin matrices\cite{easyspin}. For example, above the diagonal, the matrix
elements of the raising ladder operator appear as 
\begin{equation}
\langle n'|S^+|n\rangle=\delta_{n',n+1}\sqrt{S(S+1)-n(n+1)},
\label{spinmatrix}
\end{equation}
where $S_z|n\rangle=n|n\rangle$. The resulting spectrum consists of equidistant
energy levels at each given momentum. As mentioned above, with a different choice of 
$\alpha$, the spectrum would still be linear. The Dirac cones are robust in this sense. 
Furthermore, although the bands at a given momentum  would not necessarily be equidistant anymore, the flat band 
will still survive if certain symmetries are present, as discussed in the appendix~\ref{s:symmetry}.

For $\Delta = 0$ the wavefunction corresponding to the flat band is such that the probability of finding a particle in  even  layers is exactly zero. 
For example, in an $S=1$-trilayer,  the red plane, sandwiched between the blue
and green ones (see Fig. \ref{sp12}),  is completely blocked for the flat band
wavefunction.

The stacking we propose here is of course quite familiar. A succession of
triangular planes ABCABC... as displayed in Fig. \ref{sp12}, is just the 
face-centred cubic lattice
viewed along a [111] direction. Another stacking, ABABA..., corresponds to the
hexagonal close packed lattice structure. The hopping Hamiltonian,
Eq.~\eqref{eq:04}, is simply modified to take into account this stacking: $f
({\bf k})$ is replaced by its complex conjugate for hopping BA, CB or AC. 
For example the stacking ABCB would result in   
\begin{gather}
H_S = t \left[
\begin{array}{cccc}
0 & \alpha_{01} f ({\bf k}) &  0 &   0 \\[1ex]
\alpha^*_{01} f^*({\bf k}) & 0 &  {\alpha_{12} f ({\bf k})} &  0 \\[1ex]
0 & \alpha_{12}^* f^* ({\bf k}) & 0 & \alpha_{23}^* f^*
({\bf k}) \\[1ex] 
0 & 0 & \alpha_{23} f ({\bf k}) & 0 
\end{array}
\right],
\label{eq:flip}
\end{gather}
We will say that the chirality between the third and the fourth layer has been flipped.

The spectrum (and in particular Dirac cones and flat bands) is not affected by
this change, which affects only the phase of the matrix elements. Indeed, a unitary transformation, changing the
$i^{th}$ spinor entry $\psi_i (k) \to {\rm exp} [2i \arg f (k)] \psi_i (k)$
changes $f^* (k) \to f (k)$ in ${H}_{S,i-1, i}$. 

Around half-filling, where the continuum description applies, this corresponds to 
flipping the chirality in the Hamiltonian between adjacent layers. This change of 
chirality is at the origin of a change in the Berry curvature (detailed in
Sec. \ref{hallsec}). Note that one can successively `fix' the phases of the
off-diagonal terms to agree with a reference stacking without altering diagonal terms which may be present.

Finally, we emphasize again that for a layer separation $h < \tilde a \sqrt{2/3}$,
where $\tilde a$ is the triangular lattice constant, our Eq.~\eqref{eq:04} represents
nearest-neighbour hoppings only.

Having outlined a path towards general lattices with Dirac physics, we next
discuss some of the basic properties of such electronic systems.

\section{Density of states}

The density of states (DOS) for $\Delta=0$ is given, using the low energy Dirac Hamiltonians, by

\begin{gather}
\rho(\omega)=\sum_{{\bf p},n=-S}^S \delta(\omega-E_n({\bf p})) = \frac{A_c}{2\pi}
\frac{|\omega|}{v_F^2} \sum_{n>0}^S \frac{1}{n^2} + \nonumber \\ +\delta(\omega)\delta_{S,integer} 
\label{dos}
\end{gather}
per spin, valley and unit cell, $A_c$ being the unit cell area. The DOS remains
linear in energy, similar to graphene, but exhibits a sharp peak due to the  
flat band\cite{volovik,bercioux} for integer spin realizations. The DOS can be simplified to

\begin{equation}
\rho(\omega)=\frac{A_c}{2\pi}\frac{|\omega|}{v_F^2}\left(\frac{\pi^2}{6}-\Psi'\left(S+1\right)\right)+\delta(\omega)
\end{equation}
for integer spins, and

\begin{equation}
\rho(\omega)=\frac{A_c}{2\pi}\frac{|\omega|}{v_F^2}\left(\frac{\pi^2}{2}-\Psi'\left(S+1\right)\right)
\end{equation}
for half-integer spins, where $\Psi(x)$ is Euler's digamma function. 
Due to the momentum
integral in Eq. \eqref{dos}, these results are only valid for $|k|\ll k_c$ with
$k_c$ the cutoff, which translates to $|\omega|\ll v_Fk_c$. Note that,
for large spin $S\gg 1$, the maximal slope of the DOS right at the Fermi energy
is $\pi^2/8$ times larger than for spin-1/2 for half integer spins, and
$\pi^2/6$ times larger than for spin-1 for integer spins.
 The $S$ dependence of the slope of the DOS is shown in Fig. \ref{dosslope}, which changes very little with $S$ in the integer or half-integer sector. 
With increasing $S$, additional Dirac cones appear with increasing slope, thus with a much reduced contribution to the DOS. As opposed to that, these
high energy bands contribute more at high energies, since their bandwidth also increases with $S$.

\begin{figure}[h!]
{\includegraphics[width=6cm]{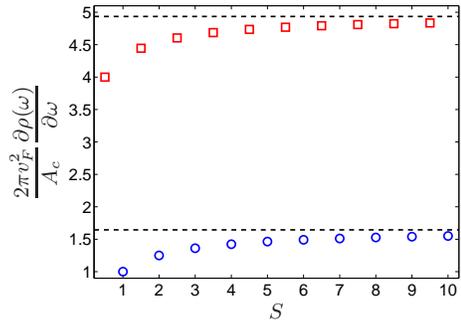}}
\caption{(Color online) The slope of the linear in energy density of states is plotted for various values of $S$, the dashed
black lines denote the asymptotic values, $\pi^2/6$ and $\pi^2/2$ for integer/half-integer spins, respectively.
\label{dosslope}}
\end{figure}

The original lattice model provides us with additional features, not captured by the low energy approximation, 
such as the presence of van Hove singularities around $\omega=n t$ with $n=-S$, $-S+1\dots S$ but $n\neq 0$ as
$\rho(\omega)\sim \ln(\omega/|n|t)$. In addition to the two peaks for graphene with $S=1/2$, increasing number of additional pairs of peaks appear in the DOS for $S>1$.

\section{Optical conductivity}

Another characteristic quantity of Dirac fermions is the optical conductivity,
which, for $S=1/2$ at half filling and $T=0$, is completely structureless and
constant. In the presence of additional Dirac bands, new interband transitions
occur. The current operator in the $x$ directions is given by
$j_x=v_FS_x$. Its equation of motion as well as those of the other spin
components are  
 
\begin{gather}
\partial_tS_x=v_Fp_yS_z,\\
\partial_tS_y=-v_Fp_xS_z,\\
\partial_tS_z=v_F\left(p_xS_y-p_yS_x\right)
\end{gather}
for a given momentum. This is easily solved for $S_x(t)$ as

\begin{gather}
S_x(t)=S_x \left[\sin^2(\varphi_{p})\cos(v_Fpt)+\cos^2(\varphi_{p})\right] +
\nonumber\\ 
+ \frac12 S_y\sin(2\varphi_{p}) \left[1-\cos(v_Fpt)\right] +
S_z\sin(\varphi_{p})\sin(v_Fpt), 
\end{gather}
where $\tan(\varphi_{p})=p_x/p_y$.

The current-current correlation function is evaluated from this as

\begin{gather}
\chi_{JJ}(t)=\sum_{n,{\bf p}}\langle S_x(t)S_x-S_xS_x(t)\rangle=\nonumber\\
=2i\sum_{n,{\bf p}}\langle S_y\rangle \sin(\varphi_{p})\sin(v_Fpt)=\nonumber\\
=2i\sum_{n,{\bf p}} np\sin^2(\varphi_{p})\sin(v_Fpt).
\end{gather}
After Fourier transformation, the optical conductivity contains two parts as

\begin{gather}
\sigma(\omega)=D\delta(\omega)+\sigma_{inter}(\omega)
\end{gather}
per electron spin and valley, and the Drude weight is

\begin{gather}
D=\frac{e^2\pi T}{h}\lfloor S+1/2\rfloor \ln\left(2\cosh\left(\frac{\mu}{2k_BT}\right)\right),
\end{gather}
which agrees with that of graphene\cite{fritz} for $S=1/2$, while
the interband part reads as

\begin{gather}
\sigma_{inter}(\omega)=-\frac{e^2\pi}{4h}\sum_{n=-S}^S nf\left(n\hbar\omega\right),
\end{gather}
where $f(x)=1/(\exp((x-\mu)/k_BT)+1)$ is the Fermi function, $\mu$ the chemical
potential, $\lfloor x \rfloor$ denotes the integer part. Since the particles  residing on the
flat band cannot propagate, their group velocity is zero, so that their
contribution vanishes to the Drude weight. This explains the integer part
function. On the other hand, they have a finite matrix element between adjacent
levels, and contribute to interband transport, which contains all allowed $2S$
processes between $2S+1$ levels.

At the Dirac point ($\mu=0$) at $T=0$, the Drude weight disappears, and the interband
conductivity reads 

\begin{gather}
\sigma_{inter}(\omega)=\frac{e^2\pi}{4h}\left(\frac{S(S+1)}{2}+
\left\{
\begin{array}{cc}
\dfrac 18 & \textmd{half-integer }S\\
0       & \textmd{integer }S
\end{array}
\right.\right).
\end{gather}
 Away from the Dirac point, $\lfloor S+1/2 \rfloor$ interband transitions
are allowed, as can be checked in Figs. \ref{optcond} and \ref{optspec}.

\begin{figure}[h!]
{\includegraphics[width=6cm,height=6cm]{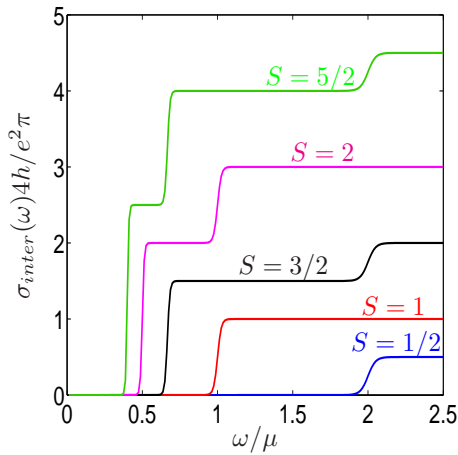}}
\caption{(Color online) The interband part of the optical conductivity for the
  spin-$S$ Dirac equation is shown for $k_BT/\mu=0.0125$ for several values of
  $S$. The number of possible interband transition is $\lfloor S+1/2 \rfloor$.
\label{optcond}}
\end{figure}

\begin{figure}[h!]
{\includegraphics[width=6cm,height=6cm]{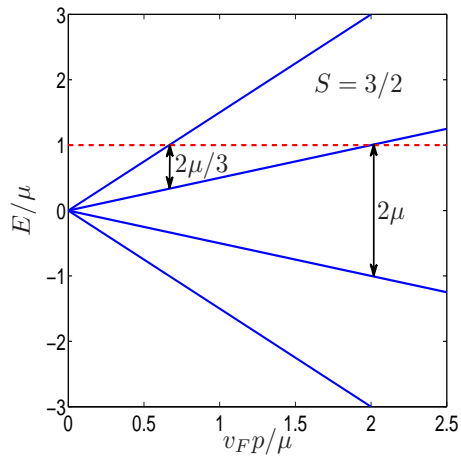}}
\caption{(Color online) The band structure of $S=3/2$, visualizing the minimal
  frequency of the allowed optical transition. As opposed to graphene,
  transitions well below $\omega=2\mu$ are possible. 
\label{optspec}}
\end{figure}

The calculated intra- and interband optical conductivities differ significantly from those in graphene. 
First, the interband part is sensitive to the number of bands and away from half-filling, several steps are possible as opposed to graphene, where only a single step is allowed.
Second, the universal value for the optical conductivity at half filling and finite frequencies is proportional to pseudospin-$S$ value, which should also affect the transparency as
\begin{gather}
\mathcal{T}=\left(1+\frac{2\pi}{c}\sigma(\omega)\right)^{-2}\approx \nonumber \\
\approx 1-\pi\alpha_{QED}\left[
S(S+1)+\left\{
\begin{array}{cc}
\dfrac 14 & \textmd{half-integer }S\\
0	& \textmd{integer }S
\end{array}
\right.\right],
\label{transeq}
\end{gather}
where the lower line is obtained upon Taylor expanding the upper line, and is only valid for $S\lesssim 2$. Here, valley and physical spin degeneracies are 
included, $\alpha_{QED}=e^2/\hbar c$ is the fine structure constant, $c$ the speed of light. 
For $S=1/2$, this reproduces the $\mathcal{T}\simeq 97.7$\% optical transparency of graphene.
Therefore, the universal value of the optical response 
immediately reveals the underlying pseudospin-$S$ structure, as shown in Fig. \ref{trans}.
Third, while the interband response takes the contribution of the flat band into account, the intraband one (Drude) is insensitive to its presence  due to the zero group velocity of the flat band.

\begin{figure}[h!]
{\includegraphics[width=6cm]{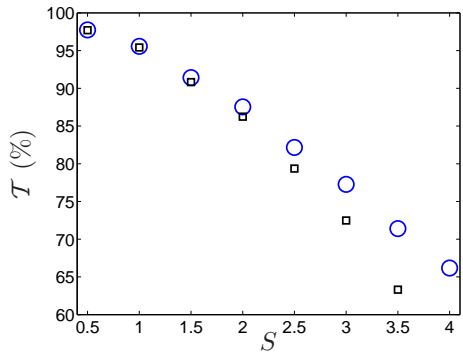}}
\caption{(Color online) The universal optical transparency is shown for half filling as a function of $S$. The blue circles denote the exact 
expression from Eq. \eqref{transeq} (upper line), while the
black squares come the approximate formula, valid for small $S$ (lower line in Eq. \eqref{transeq}).
\label{trans}}
\end{figure}

\section{Topological properties of the band structure}

\label{hallsec}
Here we discuss the topological properties our model in the absence of a gauge field. 
It will turn out that there can be topologically non-trivial grounds states. The topological 
invariant we study---the spin Chern number--- depends on the number of layers and on the band fillings. However, it does not appear to depend on 
the stacking configuration, although the Berry curvature, from which it is calculated, does.

We study the integral of the Berry curvature, where the contribution to the Berry curvature from a given band is given by 
\begin{gather}
\mathcal{B}_n= \sum_{\begin{subarray}{c}
n=-S,\\
n'\neq n
\end{subarray}}^S \frac{2\textmd{Im}\left[\langle n,{\bf
      k}|\partial_{k_x}H|n',{\bf k}\rangle \langle n',{\bf
      k}|\partial_{k_y}H|n,{\bf k}\rangle\right]}{\left[(E_{n}({\bf
      k})-E_{n'}({\bf k})\right]^2}, 
\label{chernintegral}
\end{gather}
$n$ is a band index and $|n,{\bf k}\rangle$ a single particle eigenstate of $H$ with
eigenvalue $E_{n}({\bf k})$. In the thermodynamic limit the summation over momentum turns into an integral.
 When this integral goes over a compact manifold like the Brillouin zone, one obtains a topological invariant called the first Chern number
\begin{gather}
C_n =  \int _{\mathrm{BZ}} \frac{\mathrm{d}^2 k}{2 \pi} \mathcal{B}_n \, .  
\end{gather}

A non-zero Chern number requires breaking of time reversal symmetry (TRS). This can be 
accomplished by an external magnetic field like in the integer quantum Hall effect \cite{tknn},
but also by gap terms that break TRS\cite{haldane1988}. 
Settings of the latter kind with flat bands with non-zero Chern number have been reported\cite{flat1,flat2,flat3}. 
In this section we only study  gap terms that preserve TRS, hence the Chern number is zero. However, the Chern 
number is integrated from a non-trivial Berry curvature that derives mainly from the two singularities in the Brillouin 
zone - the Dirac cones at the  $K$ and  $K'$ points. When calculating the total vorticity of a configuration of quantized vortices,
 it is usually enough to calculate the vorticity of the individual vortices as if they were isolated, and then add up the quanta 
(including the sign) to obtain the total vorticity. Similarly, it is usually enough to calculate the integral of the Berry curvature for 
 individual Dirac cones, living not on a Brillouin zone but on the infinite plane of momenta and 
then add up their contributions to obtain the same result which would originate
from  integrating the full band structure over the Brillouin zone.
 We will use both approaches and demonstrate explicitly (in Table \ref{tabc}) that they match. 

Another topological invariant is the spin Chern number. 
It is calculated from the contributions to the Chern number 
individually for the two components of the physical spin, but 
instead of adding up the two contributions to get the Chern number, 
one takes the difference to obtain the spin Chern number. 

The charge Chern number is related to a topologically quantized Hall current of charge \cite{tknn},
\begin{equation}
\sigma_{xy}=\frac{e^2}{h}\sum_n C_n\, ,
\label{sigmaxyfromchern}
\end{equation}
with the summation taken over filled bands. Likewise,  the spin Chern number can be used to determine the quantized spin-Hall conductivity.  
This is the case with the intrinsic spin-orbit coupling (SOC) of graphene \cite{kanemele1}
\begin{gather}
H_{SO}=\Delta\tau_z S_z\sigma_z,
\label{SOCS}
\end{gather}
which preserves TRS.  Here $\tau_z$ and $\sigma_z$ refers to the valley and physical spin degrees of freedom. We will generalize this SOC to arbitrary pseudospin-$S$.

In addition, the different stacking configurations lead to interesting changes in the integral of the Berry curvature in the single cone approximation, 
which calls for future research on the role of stacking order on topological properties. 


\subsection{Single pseudospin-$S$ Dirac cone}

Let us first focus on the integral of the Berry curvature 
of a single spin-$S$ Dirac equation for spinless electrons. In itself, it is usually not a topological invariant, but the topological invariants can often be understood 
in terms of the contributions for the single cones, and the latter can in the simplest case be evaluated analytically. In some cases, these non-zero contributions add up to zero, 
as must be the case for the Chern number in the time reversal symmetric  setup that we consider. However, we will also see that for the spin Chern number, another topological invariant, 
the contributions add up to an invariant that can be nonzero.    
 
For a given band in
the spin-$S$ Dirac equation, assuming a gap of the form $\Delta S_z$, 
the integral of the Berry curvature is evaluated around the $K$ point by assuming an isolated Dirac cone in the momentum plane in the expression of Eq.~\eqref{chernintegral}. 
Thus
\begin{gather}
C_n^{K}=\int\limits_0^\infty \frac{n\Delta v_F^2 pdp}{\left[(v_Fp)^2+\Delta^2\right]^{3/2}}=n\, \textmd{sign}(\Delta),
\label{intberry}
\end{gather}
where $n$ takes the allowed values of $S_z$, which also indexes the bands. 
For the $S=\frac 1 2 $ Dirac equation, this reproduces the
known result\cite{semenoff} $C_\pm=\pm\frac 12\textmd{sign}(\Delta)$ with the
upper/lower sign corresponding to the upper/lower Dirac cone. 

The contribution to the transverse
conductivity from an individual cone at the Dirac point $K$ is the sum of the above $C_n^{K}$'s from the filled bands. 
In the Dirac cone approximation a gap exists only around zero energy (between $n=-1/ 2$ and $n=1/2$ for half-integer 
spin or between the bands  $n=-1$, $n=0$ and $n=1$ for integer spin). 
Therefore, as long as the
chemical potential satisfies  $|\mu|<\Delta/2$ for half-integer and $0<|\mu|<|\Delta|$ for
integer spins,  one obtains also in the low-energy approximation a half-integer quantized transverse response
(per spin and valley)
\begin{gather}
\sigma_{xy}^{K}=\frac{e^2}{h}\sum_{n<0}
C_n^K=-\frac{e^2}{h}\frac{S(S+1)}{2}\textmd{sign}(\Delta)-\nonumber\\ 
-\frac{e^2}{h}\left\{
\begin{array}{cc}
0 & \textmd{for integer }S\\
\dfrac{\textmd{sign}(\Delta)}{8} & \textmd{for half-integer }S
\end{array}\right. \, .
\label{eq:sxy}
\end{gather}
 An increasing chemical potential  will  cut into some bands which destroys the
 half-integer quantization of $\sigma_{xy}^K$.

By choosing a stacking different from  ABCAB..., other decompositions across the bands    
can be obtained for the Berry curvature in the single cone approximation, which lead to a result different from Eq. \eqref{intberry}. 
For example, for $S=1$, i.e.\ the trilayer with ABC stacking,  
we obtain for positive $\Delta$ that  $(C_{1}^K, C_0^K, C_{-1}^K)=(1,0,-1)$ per spin
and valley.  By flipping the chirality between the second and third layer (which corresponds to an ABA stacking),
we obtain instead
$(1/2,-1,1/2)$ as integrals of the Berry curvature  for the successive
bands. (These integrals were evaluated numerically, in contrast to those related to the spin-$S$ Hamiltonians.) 
Most importantly, the 0 contribution of the flat band gets modified to
1. Various stacking patterns, whose variety grows with the number of layers,  are listed in Table \ref{tabc}. 
As we demonstrate next, the integral of the Berry curvature around a single cone can be used to determine the spin Chern number of the lattice model.

\subsection{Chern numbers on the lattice}

To obtain a topological invariant   
the full Brillouin zone (including the $K'$ point) has to be considered. For a setting with TRS, which we consider here, 
the Chern number has to be zero, even though the Berry curvature contributions around the individual cones may be nonzero, in which case they cancel.
This cancellation can be avoided by involving   the physical spin:
by generalizing the intrinsic SOC of graphene in Eq. \eqref{SOCS} 
to general spin-$S$, we can end up with non-zero {\em spin} Chern number and thus finite spin Hall conductivity.

By going back to the original lattice model and defining the full lattice version of Eq. \eqref{SOCS} following Ref. \onlinecite{fukane} as
\begin{gather}
H_{SO,lattice}=\frac{2\Delta}{3\sqrt{3}}\sigma_zS_z\left[2\sin(\sqrt{3}k_xa/2)\cos(3k_ya/2)-\right.\nonumber\\
\left.-\sin(\sqrt 3 k_xa)\right],
\label{SOlattice}
\end{gather}
which should be added to Eq. \eqref{eq:04}, 
we get the  spin Chern numbers, $C^s_n$ from Eq. \eqref{chernintegral} (with a numerical integration  over  the entire Brillouin zone) as
\begin{gather}
C^s_n=C^\uparrow_n-C^\downarrow_n=4n\, \textmd{sign}(\Delta)\, ,
\label{spinchern}
\end{gather}
where the factor $4$ comes from the valley and physical spin degrees of freedom, and $C^\sigma$ is the 
Chern number for up ($\sigma =\uparrow$) or down ($\sigma =\downarrow$) spins and
$C^\uparrow_n=-C^\downarrow_n$.

Whether the system qualifies as a spin-Hall insulator is decided\cite{fukane} by the $Z_2$ invariant $\nu$, defined by 
$\nu=\sum_n C^s_n/2\, (\textmd{mod}\, 2)=\sum_n 2n\textmd{sign}\Delta\, (\textmd{mod}\, 2)$, and summation is taken over 
filled bands.

We thus find that  integer pseudo-spins  contribute with even numbers to the sum in $\nu$,
and are topologically trivial for the lattices in Fig. \ref{sp12}, giving $\nu=0$.
By contrast, half-integers spins contribute with odd numbers to the sum, and having an 
even number of filled bands at half filling (i.e. $S=3/2$, 7/2, 11/2\dots)
adds up to an even number, thus again  $\nu=0$.
As opposed to this, half-integer spins with $S=1/2$, 5/2, 9/2\dots have an odd number of filled bands at half filling, adding up to an odd number, resulting 
in $\nu=1$ and topologically non-trivial behaviour.

\begin{table}[t!]
\centering
\begin{tabular}{|c||c|c|c|}
\hline
$S$ & stacking & \textmd{$C_n^K$ (single cone)} & $C^\uparrow_n$ \textmd{(lattice)} \\
\hline\hline
1/2 & AB & (1/2,-1/2) & (1,-1)\\
\hline
1  & ABC & (1,0,-1) & (2,0,-2) \\
1  & ABA & (1/2,-1,1/2) & (0,0,0)\\
\hline
3/2 & ABCA & (3/2,1/2,-1/2,-3/2) & (3,1,-1,-3)\\
3/2 & ABAB & (1/2,-1/2,1/2,-1/2) & (1,-1,1,-1) \\
3/2 & ABCB & (5/4,-1/4,-5/4,1/4) & (1,1,-1,-1) \\
\hline
2 & ABCAB & (2,1,0,-1,-2) & (4,2,0,-2,-4)\\
2 & ABCAC & (15/8,1/2,-3/4,-3/2,-1/8) & (2,2,0,-2,-2)\\
2 & ABCBC & (11/8,0,-3/4,0,-5/8) & (2,0,0,0,-2)\\
2 & ABCBA & (5/4,-1/2,-3/2,-1/2,5/4) & (0,0,0,0,0)\\
2 & ABABA & (1/2,-1/2,0,-1/2,1/2) & (0,0,0,0,0)\\
2 & ABACA & (0,-1,0,1,0) & (0,-2,0,2,0)\\
\hline
\end{tabular}
\caption{The integral of the Berry curvature for a single pseudospin-$S$ Dirac cone ($C_n^K$) and the spin dependent Chern number ($C^\uparrow_n$) for the lattice model are shown
for all possible,  non-equivalent stacking patterns for $S<5/2$ and $\Delta>0$. 
These are evaluated numerically for both the continuum and lattice model, and their relation can be checked using Eq. \eqref{cont-lattice}. 
Note that $C^\uparrow_n=-C^\downarrow_n$.
\label{tabc}}
\end{table}

The spin dependent Chern numbers of the lattice model can be obtained from 
the single cone results of the previous section. In the single cone approximation the spin-orbit coupling of Eq. \eqref{SOlattice} simplifies to  Eq. \eqref{SOCS}.
These single cone contributions to the Chern number satisfy 
\begin{gather}
C_n^{K}(\Delta)=-C_n^{K'}(\Delta)=C_{-n}^{K}(-\Delta)=-C_{-n}^{K'}(-\Delta).
\end{gather}
The opposite sign of the gap term for 
the two Dirac points derives from $\tau_z$ in Eq. \eqref{SOCS}.

By taking both Dirac points into account, we obtain from the single cone results the spin Chern number 
\begin{gather}
C^\uparrow_n=C_n^{K}(\Delta)+C_{n}^{K^\prime}(-\Delta)=C_n^{K}(\Delta)-C_{-n}^{K}(\Delta).
\label{cont-lattice}
\end{gather}
The result agrees with the result in Eq.~\eqref{spinchern} found from integrating over the Brillouin zone using the full band structure. 
Eq. \eqref{cont-lattice} immediately implies that the spin Chern number of the flat band is zero ($C^\uparrow_0=0$), regardless of the value of $C_0^{K}(\Delta)$.

We can also consider other stacking patterns, as we did in the single cone case.
The correspondences in Eq.~\eqref{cont-lattice} hold for arbitrary stacking patterns on the lattice.
For example, the trilayer with ABA stacking with $\Delta>0$ yields zero spin Chern numbers for all bands, unlike the $S=1$ case derived from the ABC stacking.  
However, like the ABC stacking, the ABA gives a spin Chern number that is topologically trivial. Results for other stackings are shown in Table \ref{tabc}. 
One interesting general conclusion that we can draw based on these results is that while the single cone Chern number contributions for each band are redistributed significantly 
with different stacking patterns, this does not affect the $Z_2$ topological invariant. 
This invariant  will therefore be determined only by the number of layers but not by the stacking.
We have also checked that all non-equivalent stackings for the $S=5/2$ case (not shown here) give $\nu=1$ at half filling. The invariance to changes in stacking applies also away 
from half-filling, as long as the 
chemical potential lies between the bands.   

We can also engineer  nearly flat bands with non-trivial topology, similarly to Refs. \onlinecite{flat1,flat2,flat3}:
 when $\Delta\gg t$, all bands become practically flat as $E_n(p)\approx n\Delta+nt^2|f({\bf k})|^2/2\Delta$, i.e. the hopping occurs only to second order in perturbation theory.
Therefore, it becomes possible to fill the separate bands one by one. Then, for example, 
the quarter filled $S=3/2$ case, which corresponds to a completely filled $E_{-3/2}(p)$ band, becomes topologically non-trivial with $\nu=1$ for all stackings.
Note that when $\Delta\ll t$, quarter filling in this case gives partially filled $E_{-3/2}(p)$ and $E_{-1/2}(p)$ bands.
In the same vein, the 1/3 filled $S=5/2$ lattice with flattened bands ($\Delta\gg t$) is topologically trivial with $\nu=0$. Thus, a trivial ground state can become non-trivial 
(and vice versa) when the chemical potential is lowered or increased to the next band gap.

Another observation we have made is that while the topological invariants (the Chern number and the spin Chern number) are robust with respect to the variations of $\alpha$'s 
in Eq. \eqref{eq:04}, 
the integrals of the Berry curvature for a 
single cone ($C_n^K$) are not invariant. However, for some stacking patterns, $C_n^K$ is rather insensitive to changes in  $\alpha$'s.
In particular, the single cone results for ABABA and ABACA stackings are also recovered for uniform interlayer hoppings.

We close this section with the remark  that topological invariants do not  depend  only on $S$ but also on the number of non-equivalent Dirac cones and thus on the specific form of the lattice.
 For example, 
$T_3$ and Lieb lattices\cite{qshlattice,weeks} with even and odd number of $S=1$ cones, respectively, belong to different $Z_2$ class\cite{hasankane}. 
In the presence of intrinsic SOC, the $T_3$ lattice with two inequivalent  cones, possesses the  trivial $Z_2$ index.
On the contrary, the Lieb lattice has a single cone in its band structure  and has therefore a ground state with  a non-trivial $Z_2$ invariant,
and realizes a spin-Hall insulator in the presence of SOC.


\section{Topological properties in the presence of a magnetic field}

A topological property of our lattice that turns out to depend dramatically on the stacking configuration 
is the number of zero-modes in a magnetic field.  
 In the case of a uniform magnetic field, these zero-modes are nothing 
but  $E=0$ Landau level (LL) states.  
We now show that by changing the stacking from the ABCAB... stacking to some other stacking, the $E=0$ Landau level degeneracy will increase by a multiple. 

This can immediately be seen in experiment as an 
increased step at $\mu=0$ in the steps 
of  quantized Hall conductivity  $\sigma_{xy}$  as a function of chemical potential, in a way analogous to the 
well-known examples of mono- and bilayer graphene: in the former,  all Landau levels  have 
the same degeneracy, while in the latter, the degeneracy of the $E=0$ LL is twice the one of the others 
\cite{mccannbilayer,novoselov-bilayer-iqhe}. Such degeneracies and their lifting 
play an important role, for example at integer fillings in the context of multicomponent quantum Hall ferromagnetism, 
see e.g. Ref.~\onlinecite{goerbighabil}.



The reason for this increased multiplicity thanks to restacking is rather easy to understand. Consider the gauge invariant momentum operator 
$\dirop_i ({\bf x})= -i \partial_{x_i} -A_i({\bf x})$. The existence of zero-modes relies on the fact  
one of the chiral Dirac operators $\dirop_\pm =  -i(\partial_x\pm i\partial_y) -(A_x\pm i A_y) $ has a non-trivial kernel when the net flux of ${\bf A}$ is bigger 
than one flux quantum. The number of states in the kernel is 
given by  the number of flux quanta. Depending on the sign of the total 
flux one finds non-trivial solutions either to $\dirop_+ \psi=0$ or to $\dirop_- \psi=0$, but not to both. This latter fact comes into play in an interesting way when we start 
to flip the chiralities of the hoppings between individual layers by restacking, as we will now come to.    

Like in graphene we discuss the Landau level spectrum in the linearized 
regime of the low-energy Hamiltonians, that is, in terms of the Dirac cones. 
(In the case of the full band structure we cannot write the Hamiltonian only in terms of the chiral combinations 
$\dirop_\pm$, which is necessary for the analytical 
discussion of zero-modes.)
The Dirac Hamiltonian is then written in real space and the magnetic field is  introduced by minimal coupling. 
 For the ABCA stacking, we find at the Dirac point $K$
\begin{gather}
H_\mathrm{ABCA}
=
\left[
\begin{array}{cccc}
0  & \alpha_{01}\dirop_-  & 0 & 0
\\ 
 \alpha^*_{01}\dirop_+ & 0 &  \alpha_{12} \dirop_- & 0
\\
0 &  \alpha^*_{12} \dirop_+ & 0 &  \alpha_{23} \dirop_- 
\\
0 & 0 & \alpha^*_{23} \dirop_+ & 0
\end{array}  
\right]
\label{eq:habcareal}\, .
\end{gather}
With the ABAB stacking we instead have  
\begin{gather}
H_\mathrm{ABAB}
=
\left[
\begin{array}{cccc}
0  & \alpha_{01}\dirop_-  & 0 & 0
\\ 
 \alpha^*_{01}\dirop_+ & 0 &  \alpha_{12} \dirop_+ & 0
\\
0 &  \alpha^*_{12} \dirop_- & 0 &  \alpha_{23} \dirop_- 
\\
0 & 0 & \alpha^*_{23} \dirop_+ & 0
\end{array}  
\right]\, ,
\label{eq:hababreal}
\end{gather}
that is, with the chiralities of the matrix elements relating the second and third layer flipped. Thus, by restacking it is possible to obtain several columns with only one 
chirality of Dirac operators and not both. Such columns will contribute with new zero-mode solutions and will increase the zero-mode degeneracy by a multiplicity factor.
 Assume that the flux 
is such that there are $n$ solutions  $\psi_i$ ($i=1,\ldots,n$)   
to $\dirop_+ \psi=0$ and hence no solutions to $\dirop_- \psi=0$. 
Then there are only the  $n$ zero-modes for  $H_\mathrm{ABCA}$ in Eq.~\eqref{eq:habcareal} of the form
\begin{gather}
\Psi = (\psi_i,0,0,0)^\mathrm{T}\, .
\end{gather}
$H_\mathrm{ABAB}$ in Eq.~\eqref{eq:habcareal}, on the other hand,  has the same  $n$ zero-modes, but also $n$ additional zero-modes  of the form
 \begin{gather}
\Psi= (0,0,\psi_i ,0)^\mathrm{T}\, .
\end{gather} 
which are not  solutions to $H_\mathrm{ABCA}$ because of the mixed occurrence of $\dirop_+$ and $\dirop_-$ in the third column of  $H_\mathrm{ABCA}$. Thus, the 
Hamiltonian $H_\mathrm{ABCA}$  has twice as many zero modes. For larger number of layers one has an even bigger number of  different stacking configurations to choose between, 
each with different implications for the zero-mode degeneracy. The one extreme case is given by the ``face-centred cubic'' stacking  ABCABC...  
(a configuration without flipped chiralities), which remains at the  $n$ zero-modes for an arbitrary number of layers. 
The other extreme is given by the hexagonal close-packed
stacking ABABAB... (with an alternating sequence of chiralities), 
where the number of zero modes of $2S+1$ bands 
is $n (S+1/2)$ for half-integer $S$,  and $n S$ or $n (S+1)$ (depending on the sign of the flux) for integer $S$. 
Other stackings give some intermediate multiple of $n$ zero-modes.  
A quick inspection shows that the contribution from the other Dirac point just duplicates this result for any stacking 
configuration, thus there will be a factor two due to valley degeneracy.

The above observed flexibility to increase the zero-mode degeneracy by a simple change of stacking   is in stark contrast with the situation in multilayer graphene. 
Multilayer graphene has in the simplest approximations indeed  multiple times the degeneracy of the monolayer 
Hamiltonian \cite{Falko-McCann-2006}. However, although the structure of the Hamiltonian depends sensitively on the stacking \cite{Min-MacDonald-2008}, 
the degeneracy turns out in the simplest approximation to be independent of stacking, even in the case of a non-uniform vector potential.\cite{Kailasvuori-2009b} 

These results in fact also apply for magnetic fields which are no longer uniform. While in this case, 
Landau level degeneracies  will in general  be lifted, the
$E=0$ Landau level for 2d Dirac electrons is an exception. The reason is the widely 
known general property of Dirac operators in a vector potential of arbitrary distribution which is treated by the 
Atiyah-Singer index theorem \cite{Nakahara-book}, but can be understood also on a less formal level thanks to the neat argument by Aharonov and 
Casher  \cite{Aharonov-Casher-1979}. This may imply a $E=0$ 
Landau level that is  qualitatively sharper than the other Landau levels, since only the latter are broadened by the non-uniform component of the magnetic field.  
In graphene a non-uniform component can be due to the effective magnetic field introduced by the corrugation of the graphene membrane. 
That the $E=0$ level remains relatively sharp in graphene has been observed in experiments.\cite{Giesbers-etal-PRL-2007, Martin-Feldman-Weitz-Allen-Yacoby-2010}. 
Even already for weak magnetic fields,  one can expect an increased density of states at $E=0$ due to these zero-modes. 
If the degeneracy of the zero-modes can be multiplied, as we have shown for our example, then such a peak in the density of states should grow with the same multiplicity.

Notice that the  Landau level degeneracy depends strongly on the chosen stacking of the layers,
thus also influencing the height of the zero energy peak in the density of states. This is in stark contrast to the zero field results, where
the DOS is stacking independent, as studied in Sec. III.

\section{Klein tunneling on a potential step}

Finally, we discuss Klein-tunneling of spin-$S$ Dirac
electrons. As we have seen, qualitative differences arise between half-integer
and integer spins. The transmission amplitude for spin-1/2 Dirac electrons has
been studied, in connection to graphene, in Ref. \onlinecite{klein}. The spin-1 case
and the influence of the flat band  was studied in Refs. \onlinecite{shen,apaja,urban}, and
all-angle perfect transmission was found at specific energies.
Here we
discuss Klein tunneling for the pseudospin-1 case, and focus on tunneling into the flat band, as shown in Fig.  \ref{barrier}.

But before doing so, let us discuss the general matching conditions of the wavefunction for general pseudospin-$S$. In this case,
the spinor wavefunction takes the form (omitting spatial coordinates for simplicity)
\begin{gather}
\Psi=\left(\Psi_1,\Psi_2,\dots,\Psi_{2S},\Psi_{2S+1}\right)^T.
\end{gather}
 To determine the matching conditions\cite{urban}, we integrate the eigenvalue equation of the Hamiltonian in Eq. \eqref{eq:spinham} or more generally from 
Eq. \eqref{eq:04}, after expanding it around the $K$ point, $H_S\Psi=E\Psi$
 from $x=-x_0$ to $x=x_0$ and send $x_0$ to zero.
As a result, we obtain, assuming
non-diverging scalar and vector potentials
\begin{gather}
\Psi_2(-x_0)=\Psi_2(x_0),\label{spinor2}\\
\alpha^*_{i-1,i}\Psi_i(-x_0)+\alpha_{i,i+1}\Psi_{i+2}(-x_0)=\nonumber\\
=\alpha^*_{i-1,i}\Psi_i(x_0)+\alpha_{i,i+1}\Psi_{i+2}(x_0) \textmd{ for } i=1\dots 2S-1,
\label{nomix}\\
\Psi_{2S}(-x_0)=\Psi_{2S}(x_0)\label{spinor2S}.
\end{gather}
In general, this implies that $\Psi_2$ and $\Psi_{2S}$ must be continuous, since there is only a single non-zero element in the first and last row of the Hamiltonian matrix.
In addition, Eq. \eqref{nomix}, involving other components of the spinor, contains only $\Psi_i$ and $\Psi_{i+2}$, hence there is no mixing between even and odd components.

The case of half-integer pseudospin (with even $2S+1$) implies that
 the two continuous spinor components, Eqs. \eqref{spinor2} and \eqref{spinor2S}  contain one even (2) and one odd ($2S$) index. Therefore, 
the continuity of e.g. $\Psi_2$ implies through Eq. \eqref{nomix} that of $\Psi_4$.
The continuity of $\Psi_4$ in turn implies the continuity of $\Psi_6$ and so on.
The very same procedure can be carried out for the odd components.
Therefore, each component of the spinor changes continuously, thus the whole wavefunction remains continuous across a potential barrier.

The case of integer pseudospins is different: $2S+1$ is odd, therefore only two even components ($\Psi_2$ and $\Psi_{2S}$) are required 
to change continuously (as opposed to the one even and one odd
components for half-integer $S$).
This implies that all even components must be continuous, but there are only $S$ equations for the remaining $S+1$ odd components.
Only the continuity of the linear combinations of neighbouring odd components, Eq. \eqref{nomix} with $i=1$, 3,\dots $2S-1$ is required across a barrier, but nothing
can be said about the individual components.

Similar considerations apply along the $y$ directions, in which case the wavefunction still changes continuously for half-integer $S$, while only the even components
remain continuous for integer $S$. Implicitly, this difference can be traced back to the absence or presence of a flat-band.

\begin{figure}[h!]
{\includegraphics[width=6cm,height=5cm]{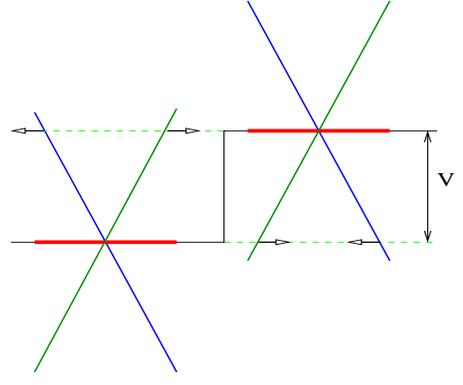}}
\caption{(Color online) A sharp potential barrier for the spin-1 Dirac
  equation, the thick red lines denote the non-propagating zero energy states,
  while the short black arrows stand for the velocity of the branches.
\label{barrier}}
\end{figure}

Here we consider the scattering of pseudospin-1 electrons on a sharp potential step of the form $V\Theta(x)$, $V>0$.
The case when the energy of the injected electron differs from $V$ has already been considered in Refs. \onlinecite{shen,urban}.
However, when the energy of the incident electron is exactly $E=V$, scattering into the non-propagating flat-band becomes possible.
 The electrons on the flat band do not possess a well defined
Fermi surface, since all particles residing on the flat band have identically
zero energy. Thus, an incident electron with $E=V$ can be scattered to any
momentum state of the flat band within the barrier. 

At normal incidence
($k_y=0$), there is perfect transmission ($T=1$), since transmission to the upper or
lower Dirac cones is possible (excluding the flat band). 

At a finite angle,
scattering to the propagating cones is forbidden by momentum conservation
($k_y$ does not change). In this special case, the wavefunction on the left and right hand side of the barrier (suppressing the $\exp(ik_yy)$ term) is given by

\begin{gather}
\Psi^L(x<0)=\frac 12 \left(\begin{array}{c}
\exp(i\varphi_k)\\
\sqrt{2}\alpha\\
\exp(-i\varphi_k)
\end{array}
\right)\exp(ik_xx)
+\nonumber\\
+\frac r2
\left(\begin{array}{c}
-\exp(-i\varphi_k)\\
\sqrt{2}\alpha\\
-\exp(i\varphi_k)
\end{array}
\right)\exp(-ik_xx),\label{eqzero1}\\
\Psi^R(x>0)=\sum_{k_x'}\frac{t(k')}{\sqrt 2}
\left(\begin{array}{c}
\exp(i\varphi_{k'})\\
0\\
-\exp(-i\varphi_{k'})
\end{array}
\right)\exp(ik_x'x)+\nonumber\\
+
a\left(\begin{array}{c}
\Theta(k_y)\\
0\\
\Theta(-k_y)
\end{array}\right)\exp(-|k_y|x),
\label{eqzero2}
\end{gather}
where $k_y$ is conserved, i.e. $k_y=k_y'$, $V=v|k|$, $\tan(\varphi_k)=k_x/k_y$,
$\tan(\varphi_{k'})=k_x'/k_y$, and the lack of Fermi surface implies that any
state on the flat band is available for transmission without any restrictions
on $k_x'$, explaining the summation over $k_x'$, $\alpha=1$. 
The last term describes an evanescent mode in the flat band.
Applying the continuity of $\Psi_2$ from Eq. \eqref{spinor2}
implies that the reflection coefficient $r=-1$, from which the reflection
probability is $R=|r^2|=1$, and $T=0$.

As far as such a stationary solution is concerned, states in the flat band to the right
of the barrier may also be occupied. However, as the group
velocity on the flat-band is zero, the transmission probability is
also zero, or in other words, the probability current is zero through the
barrier. The resulting picture thus consists of standing, non-propagating waves,
extending to both sides of the barrier: on the left, it is made of two
counterpropagating waves (in the $x$ direction), whose interference leads to a
standing wave, while on the right, the zero energy mode is non-propagating by
its very nature. Although the wavefunction to the left of the barrier is
uniquely determined by specifying energy $E=V$ and perpendicular momentum $k_y$, its
corresponding part to the right of the barrier has many degenerate versions due to the
flat band, and can thus host a large number of different states $\sim L$, even for a fixed $k_y$.

This can be made explicit as follows. Using Eq. \eqref{nomix} to connect Eqs. \eqref{eqzero1} and \eqref{eqzero2}, we get

\begin{gather}
\cos(\varphi_k)=\frac{a}{2}+\frac{i}{\sqrt 2}\sum_{k_x'}t(k')\sin(\varphi_{k'}),
\label{teq}
\end{gather}
which can have $\sim L$ distinct set of independent solutions in terms of $\{ t(k')\}$ and $a$.  Crucially, each such solutions corresponds to
zero transmission probability and perfect reflection.

Very similar considerations apply to the case of $E=0$, namely an electron in the flat-band to the left of the barrier, scattered
to propagating states to the right.

In the presence of many bands ($S>1$), tunneling between them occurs with a
greater variety, and interband tunneling is also possible between propagating
bands. However, the main difference is still expected  from the presence of
absence of a flat band (half-integer versus integer $S$), as is also reflected in the
different matching conditions.

\section{Possible experiments}

In this section, we discuss the expertimental possibilities to create optical lattices, 
which would realize the spin-$S$ Dirac equation, and
the methods to observe the characteristic physical
   quantities. This section gives a brief overview of which concrete  
   protocols have been proposed in the cold atom literature over the  
   past few years -- some of them are of course still under active    
   development. Indeed, there is a significant and ongoing
   experimental effort devoted to realise lattices with exotic band
   structures, for recent examples on triangle-based
   lattices, see Refs. \cite{kagome,struck}. Attempts to realise the
   family of lattices proposed here would form part of this endeavour.

As we have already mentioned, the lattice structure in Fig. \ref{sp12} can be regarded as 
face centred cubic lattice, which in itself -- being a Bravail lattice -- is relatively straightforward to generate.  
As a first option, this can be created by four laser
 beams at the appropriate angle, realized and discussed in detail  in Refs. \cite{grynberg,fcc1,fcc2,mei}.
Although the triangular layers would be a priori  equidistant, a setup like that of 
Ref.~\onlinecite{grynberg} already treats the lasers in one $[111]$ direction inequivalently 
from the others, so that the relative strengths for intra- and inter-plane hopping need not be equal. 
Making the former much weaker than the latter (even without adding further laser beams, relative angles 
and intensities of the beams are tunable) will then yield a bandstructure including the presence the flat 
band, as described above. 
To achieve the  chosen number layers, one can e.g.\ create an optical superlattice 
in the perpendicular direction to the layers\cite{guidoni}, or by utilizing blue-detuned light sheets to 
terminate the layered structure.
Particles are then mainly  confined to these triangular layers, whose number defines $2S+1$.

Second, one can profit from the versatility of a holographic mask, enabling arbitrary geometries, to generate 
the desired lattice structure\cite{bakr}.
Another option is to follow the steps outlined in Ref. \cite{lan} and to introduce spin dependent hopping 
amplitudes, which in 
turn also realize the desired multiple Dirac-cone structure.

In terms of observables, 
the presence of the flat band can be revealed by time-of-flight imaging, since particles residing on the flat 
band remain immobile\cite{apaja}, 
and would show up as 'missing' particles.
In addition, the number of particles (the integral of the density of states) on the lattice as a function of 
the chemical potential could be monitored, 
which a jump around zero energy for integer $S$ due to the
large degeneracy of the flat band\cite{bercioux}. The particle number per lattice site and physical spin 
behaves close to half filling as
\begin{gather}
N(\mu\simeq 0)-\frac{2S+1}{2}=\frac{\delta_{S,integer}\textmd{sign}(\mu)+\rho(\mu)\mu}{2}
\end{gather}
 where $\rho(\mu)$ is the DOS, which can be obtained by taking the numerical derivative of the experimentally 
measured particle number with respect to $\mu$.
The DOS can also conveniently be measured by rf spectroscopy, which directly probes the momentum integrated 
spectral function, i.e. the DOS\cite{mahan}.
The momentum resolved Raman spectroscopy can also be used for the same purpose.

The density-density correlation function, which is readily related to the optical conductivity\cite{mahan} can 
be investigated by shot noise measurement,
while the optical conductivity can directly be probed by the amplitude or phase modulation of the optical 
lattice\cite{tokuno}. Thanks to the modulation, 
the energy absorption rate or the doublon production rate 
turns out to be directly proportional to $\sigma(\omega)$.

To probe the spin-Hall effect, an effective electric field should be applied by
tilting the lattice along one direction, and the detection of the spin current accumulation
through separate imaging of the two different spin components\cite{shexp1} could reveal the 
quantization of the spin-Hall conductivity, stemming from the underlying topology of the band structure.

Work in progress on realising artificial gauge fields holds the promise
to probe the Landau level degeneracy. These gauge fields mimic the effect of a real vector potential, thus 
leading to the formation
of Landau levels. The enhanced degeneracy of the zero energy level should be revealed by time-of-flight imaging, 
as discussed above, or by rf spectroscopy. 
In addition, the Hall current  can be made visible by driving the system out of equilibrium by suddenly changing 
the trapping potential, and measuring the Hall current\cite{palmer}.

Additional methods for detecting topological properties, such as quantized Hall conductivity\cite{oktel} and chiral edge 
states\cite{goldman} have been discussed recently in the literature and could
be generalized to our lattice setup.

Klein tunneling is expected to be observed in the presence of smooth potential 
barrier, achievable by an accelerated optical lattice potential\cite{dahan} or by simply tilting the 
lattice\cite{phystoday}. 
Both methods would give rise to an additional potential term, varying linearly in one direction as $V({\bf r})\sim x$.
A sharp potential barrier is also available using the appropriate holographic masks.
The characteristics of Klein tunneling (i.e. perfect transmission at given angles) should show up in the 
measured momentum distribution.

Finally, we mention in passing that photonic crystals allow the realization of the appropriate lattice 
geometry\cite{escuti}, sketched in Fig. \ref{sp12}.
The nature of the edge states can be probed similarly to Ref. \cite{photonic}, together with the characteristics 
of Klein tunneling.


\section{Conclusions}

In conclusion, we have studied the lattice generalization of the spin-1/2 Dirac
equation of graphene to arbitrary spin. The main difference arises between
integer and half-integer spins, the former possessing a flat band, which is
absent for graphene (corresponding to $S=1/2$). As a result, the density of
states and the optical conductivity are modified, and the topological
properties are also enriched. We would like to reemphasize the following points.

First, even in the absence of a perfect $\bf S\cdot p$ Hamiltonian, the above
multi-cone picture can survive with a different ratio of the opening angles
between the cones.  

Second, a flat band is expected under general conditions in the gapless case: any lattice with an
odd number of layers and with chiral symmetry is expected to have a flat band. In the gapped case 
the flat band is less general, but can still be realised in some natural settings, among  the case of equidistant layers with regular stacking.  

Third, some topological properties depend sensitively on the stacking configuration. We have established 
this for the Aharonov-Casher zero-modes in a random magnetic field. Also the Chern numbers of individual bands
are strongly stacking dependent.

Fourth, while the wavefunction remains continuous across a potential barrier for half-integer pseudospin, only its even components remain continuous
for integer $S$.

Last, we would like to emphasize that the lattice structure in Fig. \ref{sp12} simply corresponds to
a four-layer slab of the face-centred cubic Bravais lattice squeezed in the
[111] direction.

\begin{acknowledgments}

We thank J\'anos Asb\'oth, Hanspeter Buechler, J\'ozsef Cserti, Graham Kells, Ferenc Simon, Shivaji Sondhi and
Attila Virosztek for stimulating
discussions and comments, Dario Bercioux for pointing out an error in the previous version and for bringing
Ref. \onlinecite{urban} to our attention.
Support by the Hungarian Scientific Research Fund
No. K72613, K73361, CNK80991,
New Sz\'echenyi Plan Nr.
   T\'{A}MOP-4.2.1/B-09/1/KMR-2010-0002 and by the Bolyai
program of the Hungarian Academy of Sciences is acknowledged.

\end{acknowledgments}

\appendix
\section{Symmetry conditions for  flat bands}
\label{s:symmetry}
In this appendix we discuss in further detail  the different symmetry properties both  of the  
full Hamiltonian in Eq.~\eqref{eq:04} as well as of  its low-energy expansion around the Dirac points of which the spin $S$ Hamiltonian
 Eq.  \eqref{eq:spinham} is a special case. We will show that the existence of the flat band in the different settings 
can be understood in terms of symmetries and   that the symmetry conditions in the different settings are of different level of generality, 
that is, of different level of robustness to variations in the parameters of the model. 

In the gapless case the 
flat band is protected by a very general symmetry---the chiral symmetry---which is made possible by the  bipartite nature of the lattice, 
and is present  under the  considered nearest neighbor hopping, which respects this bipartitness (bipartiteness will be explained below).  
The flat band is insensitive to arbitrary variations of the parameters $\alpha_{ij}$, for example due to small misalignments of the lattice, 
as this does not change the bipartite structure of the Hamiltonian. 

The situation is changed when one introduces any diagonal terms, 
for example due to intralayer hopping or in the form of   a gap term like for example $\Delta S_z$. Such terms  do not respect  the bipartite 
structure and leads  to the violation of the the chiral symmetry.
However, less general symmetries protecting flat bands can still be at play. 
The flat band will no longer be robust to any small arbitrary variations in the parameters. 
However, it will still be there for a rather wide class of parameter configurations, including fortunately some rather natural configurations like equidistant layers. 
We will now discuss these symmetry issues in further detail, starting with the chiral symmetry.       

A lattice is bipartite if the sites can be collected into two partitions $A$ and $B$ and the Hamiltonian only contains nonzero 
matrix elements between the partitions but not within each partition (which is the case of the nearest neighbor hopping on the honeycomb lattice, which gives 
the free massless Dirac Hamiltonian).  In our multilayer generalization the partitions are the odd and the even layers, respectively. Chiral symmetry  
is present as the only terms in the Hamiltonian are the hoppings  between adjacent layers. In contrast, hopping {\it within} the layers of triangular 
lattices would be one source of diagonal terms in the Hamiltonian Eq.~\ref{eq:04}. A transverse electric field would be another source.   
Such diagonal terms break the chiral symmetry.  

We are now going to discuss on a more mathematical level the conjugation symmetry properties of the model Hamiltonian
\begin{widetext}
 \begin{gather}
H_S = t \left[
\begin{array}{cccc}
\Delta_0 & \alpha_{01} f ({\bf k}) &  0 & \\[1ex]
\alpha^*_{01} f^*({\bf k}) & \Delta_1 &  {\alpha_{12} f ({\bf k})}_{\ddots} &  0 \\[1ex]
0 & \alpha^*_{12} f^* ({\bf k})_{\ddots} & 0_{\ddots} & \alpha_{2S-1, 2S} f
({\bf k}) \\[1ex] 
& 0 & \alpha^*_{2S-1, 2S} f^* ({\bf k}) & \Delta_S 
\end{array}
\right] \, .
\label{eq:hamsym}
\end{gather}
\end{widetext}
 When the gap terms are zero $\Delta_i=0$, the Hamiltonian possesses a chiral symmetry (CS):  
$\Sigma H({\bf k})\Sigma^\dagger =-H({\bf k})$, where $\Sigma$ is a unitary matrix. The CS conjugates the spectrum. 
In our case we have $\Sigma=\textrm{diag} (1,-1,1,-1,\ldots)$ generalizing  $\Sigma=\sigma_z$ that conjugates  
the gapless  $S=\tfrac{1}{2}$ Dirac spectrum. In the case of an odd number of bands the CS has to map the middlemost band  onto itself. 
Furthermore, since the CS conjugates the spectrum 
at each $\bf k$ separately, each (crystal) momentum eigenstate in the middle band has to map onto itself.     
Chiral symmetry guarantees thus  a zero-energy state at each ${\bf k}$---a flat band.
 This should be contrasted with  particle-hole symmetry (PHS)   
$U_\mathrm{PH}  H^*({\bf k}) U_\mathrm{PH}^\dagger = -H(-{\bf k}) $, also possessed by  the same Hamiltonian.
PHS conjugates the band structure, however,  does not do so at each ${\bf k}$ independently wherefore PHS alone is not enough to guarantee a 
flat band. \footnote{One could imagine a  single band $\sim \sin k_x+\sin k_y $. It is not flat, but has the self-conjugate property upon ${\bf k} \rightarrow -{\bf k}$.}
 
The chiral symmetry is not only independent  of the precise  values of the parameters $\alpha_{ij}$'s, but also of the stacking.  
As the chiral symmetry only relies on bipartiteness, it is also independent to changes in the stacking with 
the consequent individual flipping of chiralities in the tunneling between the involved layers, as for 
 the Hamiltonian Eq. \eqref{eq:flip}.

For the gapped case  the chiral symmetry is lost and there are no general symmetries to guarantee a flat band. 
Nonetheless, the following conditions are still sufficient for the appearance of a flat band
 in the general case with a nonzero gap. The diagonal gap term  
$\mathrm{diag}(\Delta_0, \Delta_1, \ldots,  \Delta_{2S})$ must satisfy the anti-symmetric property $\Delta_{2S}=-\Delta_0$, $\Delta_{2S-1}=-\Delta_1$ etc. 
At the same time, the off-diagonal elements with the $\alpha$ parameters must satisfy a symmetric property in their moduli:   $|\alpha_{2S-1,S}|=|\alpha_{01}|$,   
$|\alpha_{2S-2,S-1}|=|\alpha_{12}|$ etc., while the phases can be chosen arbitrarily. For this reason, 
the flipping of chiralities between layers does not change anything since $|f^*|=|f|$. 

Even  though it is difficult to interpret these general conditions in terms of symmetries,
these can be connected, in some special cases, to a symmetry property 
specific to the Dirac cone, that is, they are present only in  the low-energy approximation for the system close to half-filling. 
For example, if one imposes the stronger condition    $\alpha_{2S-1,S}=\alpha_{01}$,  
$\alpha_{2S-2,S-1}=\alpha_{12}$, ...  and a symmetric configuration of flipped chiralities,  
then the flat bands can be seen as a consequence of the emergent  conjugation property  $Y H_S^*({\bf  p}) Y^\dagger= -H_S ({\bf p})$, 
with ${\bf p}$ being the small momentum with respect to a Dirac point and   
with $Y$ generalizing $\sigma_y$ as a matrix with the alternating pattern  $(-i,i,-i,i, \ldots)$ on the anti-diagonal of the matrix  
(i.e., the diagonal joining bottom left and top right entries).  It formally looks like a PH symmetry combined with inversion symmetry $H(-{\bf p})=H({\bf p})$, 
however, with momenta not inverted around ${\bf k}=0$ as the physical PH symmetry requires but around the Dirac point ${\bf p}={\bf k}-{\bf k}_+=0$. 

As mentioned in the beginning of the section, some of the specific parameter configurations are rather natural. For equidistant layers,
 the gap parameter is $\Delta S_z$ and can be created by the electrostatic potential of a uniform transverse electric field. 
The  $\alpha$'s would all have the same modulus for equidistant layers and therefore also fulfill the above conditions, 
although they would not correspond to the values pertinent to  $S_x$ and $S_y$. Thus, we conclude that equidistant layers for 
certain symmetric stackings are enough to guarantee flat bands also in the presence of a gap term generated by a transverse uniform electric field.



\end{document}